\documentclass[]{spie}  

\usepackage{amsmath,amsfonts,amssymb}
\usepackage{graphicx}
\usepackage[colorlinks=true, allcolors=blue]{hyperref}

\title{SUIM project: measuring the upper atmosphere from the ISS by observations of the CXB transmitted \\ through the Earth rim}

\author[a]{Kumiko K. Nobukawa}
\author[b]{Ayaki Takeda}
\author[c]{Satoru Katsuda}
\author[d]{Takeshi G. Tsuru}
\author[e]{Kazuhiro Nakazawa}
\author[b]{Koji Mori}
\author[d]{Hiroyuki Uchida}
\author[f]{Masayoshi Nobukawa}
\author[b]{Eisuke Kurogi}
\author[a]{Takumi Kishimoto}
\author[a]{Reo Matsui}
\author[a]{Yuma  Aoki}
\author[a]{Yamato Ito}
\author[a]{Satoru Kuwano}
\author[b]{Tomitaka Tanaka}
\author[g]{Mizuki Uenomachi}
\author[d]{Masamune Matsuda}
\author[c]{Takaya Yamawaki}
\author[h]{Takayoshi Kohmura}
\affil[a]{Faculty of Science and Engineering, Kindai University, 3-4-1 Kowakae, Higashi-Osaka 577-8502, Japan}
\affil[b]{Department of Applied Physics and Electronic Engineering, Faculty of Engineering, University of Miyazaki,1-1 Gakuen Kibanadai Nishi, Miyazaki, Miyazaki 889-2192, Japan}
\affil[c]{Graduate School of Science and Engineering, Saitama University, 255 Shimo-Ohkubo, Sakura, Saitama 338-8570, Japan}
\affil[d]{Department of Physics, Faculty of Science, Kyoto University, Kitashirakawa Oiwake-cho, Sakyo-ku, Kyoto 606-8502, Japan}
\affil[e]{Kobayashi-Maskawa Institute for the Origin of Particles and the Universe, Nagoya University, Furo-cho, Chikusa-ku, Nagoya, Aichi 464-8601, Japan}
\affil[f]{Faculty of Education, Nara University of Education, Takabatake-cho, Nara, Nara 630-8528, Japan}
\affil[g]{Institute of Innovative Research, Tokyo Institute of Technology, 2-12-1 Ookayama, Meguro-ku, Tokyo 152-8550, Japan}
\affil[h]{Department of Physics, Tokyo University of Science, 2641 Yamazaki, Noda, Chiba 270-8510, Japan}

\authorinfo{Further author information: (Send correspondence to K.K.N.)\\K.K.N.: kumiko@phys.kindai.ac.jp}

\begin{document} 
\maketitle

\begin{abstract}
The upper atmosphere at the altitude of 60--110 km, the mesosphere and lower thermosphere (MLT), has the least observational data of all atmospheres due to the difficulties of in-situ observations. Previous studies demonstrated that atmospheric occultation of cosmic X-ray sources is an effective technique to investigate the MLT. Aiming to measure the atmospheric density of the MLT continuously, we are developing an X-ray camera, ``Soipix for observing Upper atmosphere as Iss experiment Mission (SUIM)'',  dedicated to atmospheric observations. SUIM will be installed on the exposed area of the International Space Station (ISS) and face the ram direction of the ISS to point toward the Earth rim. Observing the cosmic X-ray background (CXB) transmitted through the atmosphere,  we will measure the absorption column density via spectroscopy and thus obtain the density of the upper atmosphere. The X-ray camera is composed of a slit collimator and two X-ray SOI-CMOS pixel sensors (SOIPIX), and will stand on its own and make observations, controlled by a CPU-embedded FPGA ``Zynq''. We plan to install the SUIM payload on the ISS in 2025 during the solar maximum. In this paper, we report the overview and the development status of this project.
\end{abstract}


\section{INTRODUCTION}
\label{sec:intro}  
Observations of the mesosphere and lower thermosphere (MLT: 80--110~km) are important for understanding climate change, global warming, and space weather. For example, the mesosphere and thermosphere temperatures will cool due to increase of CO$_2$, which indicates that global change will occur in the MLT [\citenum{Roble1989}]. However, the MLT has the least observational data of all atmospheres due to the difficulties of in-situ observations by satellites (altitudes above 300~km) or balloons (altitudes below 50~km). 
Determan et al. (2007) [\citenum{Determan2007}] demonstrated that occultation of cosmic X-ray sources is an effective technique to measure the atmospheric density of the MLT.  
X-rays penetrating through the atmosphere are photoelectrically absorbed by inner K-shell and L-shell electrons in atoms (including atoms within molecules) and decline in intensity, especially in the low-energy band. Spectral fitting allows us to measure the column densities of the atmosphere, which indicates atomic number densities integrated along the line of sight.
Applying the technique, Katsuda et al. (2021, 2023) [\citenum{Katsuda2021}, \citenum{Katsuda2023}] and Yu et al. (2022) [\citenum{Yu2022}] analyzed the data obtained by X-ray astronomy satellites during the atmospheric occultation of  the Crab Nebula and investigated a long-term trend of the atmosphere density. 
But the data obtained by X-ray astronomy satellites were sparse. They observe atmospheric occultations for typically only one minutes per orbit just before/after occultations of the solid Earth. Furthermore, the satellites observe the Crab Nebula only once or twice a year for the aim of calibration.  

For continuously monitoring atmospheric occultations, we are developing a new instrument dedicated to atmospheric observations, ``Soipix for observing Upper atmosphere as Iss experiment Mission (SUIM)''.  This instrument is planned to be mounted on a platform on the exterior of the International Space Station (ISS), the Materials International Space Station Experiment (MISSE), for six months in 2025 during the solar maximum. 
Utilizing the ISS platform provides us two advantages: we can shorten the development period since no bus system development is necessary, and the instrument mounted on the ISS can uninterruptedly observe the atmosphere with the same attitude.   

Figure~1a represents a schematic view of the SUIM observations. The SUIM payload placed facing the ram direction, will observe the atmospheric absorption of the cosmic X-ray background (CXB) in the energy range of 3--10~keV to measure atmospheric neutral density at each altitude in the 60--150~km range. Neutral density of the atmosphere will be obtained by X-ray spectroscopy. 
In this paper, we report the overview of the SUIM project and the development status.

\begin{figure} [hpbt]
   \begin{center}
   \begin{tabular}{c} 
   \includegraphics[width=12cm]{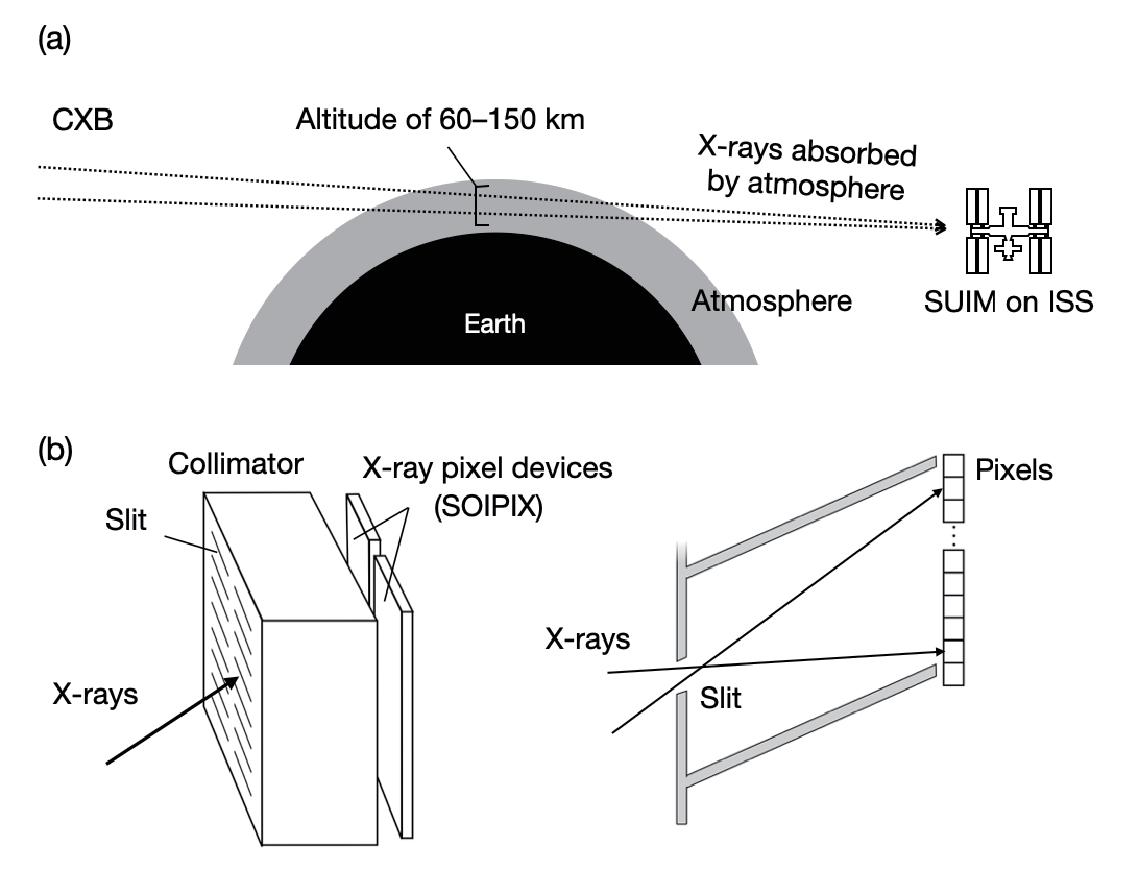}
	\end{tabular}
	\end{center}
   \caption[example] 
   { \label{fig:example} 
   (a) Schematic view of atmospheric observations by SUIM. (b) Geometry of the collimator and two X-ray pixel detectors in SUIM (left) and the cross-sectional view perpendicular to a slit (right). X-rays with different incident angles arrive at different pixels.
   }
   \end{figure}

\section{SUIM Overview}
The X-ray imaging sensors we utilize are Silicon-On-Insulator (SOI) Pixel detectors (SOIPIX), which we have developed based on silicon-on-insulator (SOI) technology [\citenum{Arai2011}]. The SOIPIX provides both high spectral power with the readout noise of $\sim$10~e$^{-}$ and high X-ray sensitivity achieving the depletion layer thickness of $300$~$\mu$m. The latter enables us to detect X-rays above 10~keV and then to measure the atmospheric density at altitudes below 90~km. 
Since each pixel of SOIPIX has its own trigger circuit, which generates a trigger signal when an X-ray signal exceeds the threshold voltage of the pixel. Hence we can read only the pixels where an X-ray photon incident  (event-driven readout).  This feature is distinct from X-ray CCDs, where signals from all pixels are read out every time (frame readout). 
SUIM will utilize two large SOIPIX devices, XRPIX-11, each of which has 592 (vertical)$\times$300 (horizontal) pixels and a size of 21.9~mm$\times$13.2~mm as the effective area with the pixel size of $36~\mu$m$\times36~\mu$m. SUIM will be the first mission to operate SOIPIX in space.  

The SUIM payload has a slit collimator placed in front of two X-ray imaging sensors as shown in figure1b.  The collimator has sixteen slits aligned parallel to the Earth's horizon. 
Eight slits are for each X-ray detector, and each slit is for 73 $\times$ 300 pixels on each device. The field of view (FOV) per each pixel, that is a spatial resolution of SUIM, is 0.3~deg (vertical) $\times$ 22~deg (horizontal).  The horizontal FOV is determined so that it is not obstructed by the ISS itself.
The vertical FOV is equivalent to the spatial resolution of $\sim15$~km in the upper atmosphere. The vertical FOV of the whole detectors is 2.2~deg, which is equivalent to the altitude range of $\sim60$--150~km. By accumulating X-ray photons along the horizontal pixels, we will measure the atmospheric density every 15~km in the altitude range of 60--150~km. 
Kishimoto et al. (2024) [\citenum{Kishimoto2024}] conducted a feasibility study of the SUIM observation and estimated the photon statistics to be $10^3$~photons (excluding the non X-ray background) for every 15 km during the six-month exposure,  which is enough to evaluate the atmospheric density.
Since the MISSE facing the ISS ram direction will point 8 degrees above the ISS forward, the collimator plates are angled so that SUIM is pointing $\sim24$~degrees downward from the facing direction of the MISSE to observe the Earth rim (see the right panel of figure~1b). 

Figure~2 represents the configuration of the SUIM payload. 
SUIM has six electronic boards: the insulated DCDC board, the control board, the high-voltage (HV) board, the camera board, the camera HV board, and the Peltier board. 
The 28V voltage from the ISS is supplied to the control board via an insulated DC-DC converter installed on the insulated DCDC board. 
The two imaging sensors are mounted on a camera board and are controlled by the CPU-embedded FPGA ``Zynq'' (Xilinx) on the control board. The signals from the sensors are transferred to an analog-to-digital converter (ADC) on the camera board and sent to the FPGA, and the control signals are sent from the FPGA to the sensors through the digital-to-analog converter (DAC).  The control board contains power supplies, power monitors, and temperature monitors.
A DC-DC boost converter module installed on the HV board can boost the voltage from the insulated DCDC board up to $-300$~V, which is applied to the X-ray sensors via the camera HV board as the back-bias voltage.  To decline the dark current, the temperature of the X-ray sensors is controlled by a Peltier device on the Peltier board to keep below 0$^{\circ}$C. The Peltier device is controlled by the IoT board computer ``SPRESENSE main board'' (SONY) installed on the HV board.  
To verify the FOV, ``SPRESENSE High Dynamic Range camera board'' with an optical CMOS camera will be installed on SUIM, which is also controlled by the IoT board computer. 
Software of the FPGA and IoT board computer will be designed so that the payload stands on its own and makes observations without commands from the ground.
The total power is expected to be 50~W. 
The payload housing is made of Aluminum and box-shaped, which also functions as an electromagnetic shield.
The housing size is 232~mm$\times$120~mm$\times$94~mm, and the total mass will be up to 5~kg. 
The housing and collimator will be manufactured as one unit by EXEDY Corporation.  
 
 We will utilize the MISSE Science Carriers (MSC) for the project.  The MSC has exposure decks, and the SUIM payload will be integrated into the MSC under the exposure deck.
The deck will have two exposure windows for the collimator and optical CMOS camera of SUIM.

\begin{figure} [ht]
   \begin{center}
   \begin{tabular}{c} 
   \includegraphics[width=15.5cm]{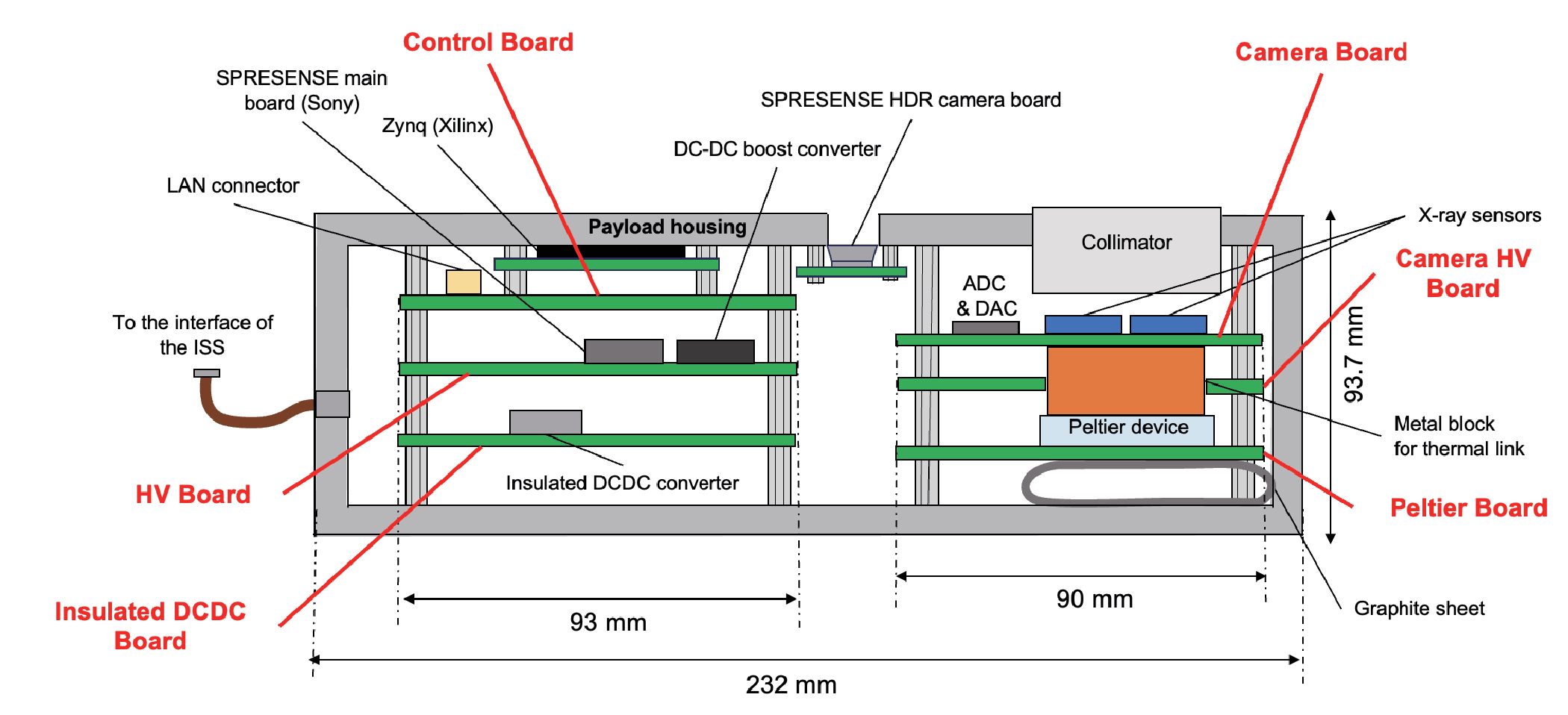}
	\end{tabular}
	\end{center}
   \caption[example] 
   { \label{fig:example} 
   Configuration of the SUIM payload. Sizing ratio are inaccurate. Cables are not described for brevity.
   }
   \end{figure}

\section{SUIM Development status}
We began the conceptual design of SUIM in April 2022, and have developed the instrument since April 2023.   
The engineering model production of the control board was completed in March 2024 and debugging work is in progress. The camera board, the housing, and the collimator are currently in production.  
The radiation test of IC components on the control board and camera board was conducted in January 2024 at the National Institutes for Quantum Science and Technology. The samples were irradiated with gamma rays from $^{60}$Co. The total irradiation dose was 1~Gy, equivalent to about two years on the ISS orbit [\citenum{Kodaira2021}]. We confirmed no problem with the components' function after irradiation.

Production of the flight model of the SUIM instrument will be completed by the third quarter of 2024 and will undergo ground testing at the end of 2024. If on schedule, the payload will be brought to the U.S. in the spring of 2025 and launched in the summer of 2025. After six months of exposure, the payload will be returned to the Earth in 2026.    

\acknowledgments 
This work was supported by Grants-in-Aid for Scientific Research from the Ministry of Education, Culture, Sports, Science and Technology (MEXT) of Japan, No. 23H00151, 23K22540, 22H01269 and MEXT Coordination Funds for Promoting Aerospace Utilization, Japan Grant Number JPJ000959. This work was also achieved by Murata Science and Education Foundation, and Mitsubishi Foundation. The authors are also grateful to EXEDY Corporation for collaboration on this work. 

\bibliography{report} 
\bibliographystyle{spiebib} 

\end{document}